\documentclass[aps,prd,superscriptaddress,twoside,twocolumn,nofootinbib,preprintnumbers]{revtex4}

\usepackage{amsmath}
\usepackage{graphicx}

\begin{document}
\preprint{\parbox[b]{1in}{ \hbox{\tt KEK-TH-1834} }}

\title{2 TeV Walking Technirho at LHC?}

\author{Hidenori S. Fukano}
\thanks{\tt fukano@kmi.nagoya-u.ac.jp}
      \affiliation{ Kobayashi-Maskawa Institute for the Origin of Particles and 
the Universe (KMI) \\ 
 Nagoya University, Nagoya 464-8602, Japan.}
\author{Masafumi Kurachi} \thanks{\tt kurachi@post.kek.jp}
      \affiliation{Institute of Particle and Nuclear Studies,
High Energy Accelerator Research Organization (KEK), Tsukuba 305-0801, Japan}
\author{Shinya Matsuzaki}\thanks{\tt synya@hken.phys.nagoya-u.ac.jp}
      \affiliation{ Institute for Advanced Research, Nagoya University, Nagoya 464-8602, Japan.}
      \affiliation{ Department of Physics, Nagoya University, Nagoya 464-8602, Japan.} 
\author{Koji Terashi}
\thanks{\tt Koji.Terashi@cern.ch}
\affiliation{The University of Tokyo, International Center for Elementary Particle Physics and Department of Physics, 
7-3-1 Hongo, Bunkyo-ku, JP - Tokyo 113-0033, Japan.}     
\author{{Koichi Yamawaki}} \thanks{
      {\tt yamawaki@kmi.nagoya-u.ac.jp}}
      \affiliation{ Kobayashi-Maskawa Institute for the Origin of Particles and the Universe (KMI) \\ 
 Nagoya University, Nagoya 464-8602, Japan.}

\date{\today}

\begin{abstract}
The ATLAS collaboration has recently reported an excess of about 2.5 $\sigma$  global significance at around 2 TeV 
in the diboson channel with the boson-tagged fat dijets, which may imply a new resonance beyond the standard model.   
We provide a possible explanation of 
the excess as the isospin-triplet technivector mesons 
(technirhos, denoted as $\rho_\Pi^{\pm,3}$) of the walking technicolor in the 
case of the one-family model as a benchmark. 
As the effective theory for the walking technicolor at the 
scales relevant to the LHC experiment, we take a scale-invariant version of the hidden local symmetry 
model  so constructed as to accommodate technipions, technivector mesons, and the technidilaton
in such a way that the model respects spontaneously broken chiral and scale symmetries of the 
underlying walking technicolor. 
In particular, the technidilaton, a (pseudo) Nambu-Goldstone boson of the (approximate) scale symmetry predicted in the walking technicolor, has been shown to
be successfully identified with the 125 GeV Higgs.  
Currently available LHC limits on those technihadrons are used to    
fix the couplings of technivector mesons to the standard-model fermions and weak gauge bosons.    
We find that the technirho's 
are mainly produced through the Drell-Yan process 
and predominantly decay to the dibosons, 
which accounts for the currently reported excess at around 2 TeV.  
The consistency with the electroweak precision test and 
other possible discovery channels of 
the 2 TeV technirhos 
are also addressed.

\end{abstract}
\maketitle


{\it Introduction.} --  
The key particle responsible for the origin of the mass, a Higgs boson,  was discovered at the LHC in 2012~\cite{Aad:2012tfa}. 
However, the dynamical origin of the Higgs 
and hence of the electroweak (EW) symmetry breaking is yet to be elucidated, suggesting existence of a theory beyond the Standard Model (SM)
to be explored at the LHC Run-II. 

One of the candidates for the theory beyond the SM 
to account for  
the dynamical origin of the EW symmetry breaking 
is the walking technicolor based on the approximately scale-invariant gauge dynamics, 
which has a large anomalous dimension $\gamma_m \simeq 1$, and the technidilaton, a pseudo Nambu-Goldstone (NG) boson of the approximate scale symmetry, as 
a light composite Higgs~\cite{Yamawaki:1985zg}. 
In this theory the EW symmetry breaking is triggered by the technifermion condensation $\langle \bar{F}F  \rangle$, like 
the quark condensation $\langle \bar{q}q \rangle$ in QCD, and 
the technidilaton arises as a composite flavor-singlet scalar meson formed 
by the $\bar{F}F$-bound state, directly linking to the dynamical breaking mechanism of the 
EW symmetry.     In a particular walking technicolor model, the ``one-family model'' with $N_F=8$ technifermions, the technidilaton   
successfully accounts for the 125 GeV Higgs at LHC~\cite{Matsuzaki:2012gd,Matsuzaki:2012vc,Matsuzaki:2012mk,Matsuzaki:2012xx,Shinya:scgt15}.

The one-family model, originally proposed  
 as a naive scale-up version of QCD~\cite{Farhi:1979zx,Farhi:1980xs}, in fact turned out to be
a promising candidate for the walking technicolor in the new context of  large $N_F$ gauge theory with $SU(N_C)$ gauge group~\cite{Appelquist:1996dq} 
through the conformal phase transition \cite{Miransky:1996pd} characterized by the Miransky scaling which is the crucial ingredient of 
the walking technicolor~\cite{Yamawaki:1985zg}:  the mass and the couplings of the
 technidilaton  
were calculated 
through both the ladder calculations and the holographic model, with the best fit to the 125 GeV Higgs for $N_F=8$ and $N_{C}=4$~\cite{Matsuzaki:2012gd,Matsuzaki:2012vc,Matsuzaki:2012mk,Matsuzaki:2012xx}
\footnote{
In the ladder calculations, the light technidilaton 
can naturally be realized in the large $N_{C}$ and $N_F$ limit, with fixed $N_F/N_{C}\, (\gg 1)$ (``anti-Veneziano limit''), in a sense similar to the
$\eta^\prime$ meson in QCD as a pseudo NG boson in the same limit, with $N_F/N_{C} \, (\ll 1)$ (Veneziano limit), both having the vanishing mass limit
though not exact massless point. See discussions in Refs.~\cite{Shinya:scgt15,Matsuzaki:2015sya}. The holographic model \cite{Matsuzaki:2012xx} gives even more reduction of the technidilaton mass due to the
large technigluon condensate which is missing in the ladder approximation.
}. 
Among other walking technicolor models, the one-family model is special in the sense that it is the most straightforward and natural model setting for the extended technicolor (ETC) as a standard way to give masses to the quarks and leptons. The value $N_{C}=4$ is further selected as an interesting  model setting within the ETC framework~\cite{Kurachi:2015bva}.
Furthermore, recent lattice studies~\cite{Aoki:2013xza, Appelquist:2014zsa, Hasenfratz:2014rna} suggest that $N_F=8$ QCD 
has desired walking signals of the condensate $\langle \bar{F}F  \rangle$ with an approximate scale symmetry and a large anomalous dimension $\gamma_m \simeq 1$. 
Moreover, a light flavor-singlet scalar meson as a candidate for the technidilaton  has been
discovered on the lattice for the $N_F=8$ QCD~\cite{Aoki:2014oha}~\footnote{
A similar 
light flavor-singlet scalar meson on the lattice has also been found in the $N_F=12$ QCD \cite{Aoki:2013zsa,Fodor:2014pqa,Brower:2014ita}. Although 
many lattice results indicate that the $N_F=12$ is in the conformal window where the chiral and scale symmetries 
are not spontaneously broken and hence no dilaton exists, such a light scalar in $N_F=12$ QCD may be a generic feature of  the conformal (scale-symmetric)  
dynamics common to $N_F=8$ QCD.
}.

Besides the technidilaton at 125 GeV, the one-family model  of the walking technicolor  
predicts a rich composite spectra in several-TeV region to be seen at LHC Run II: 
the model consists of  techniquarks (3-colored doublets)  and technileptons (1-noncolored doublet),
having the same SM quantum numbers as those of 
the one-generation of the SM fermions, and hence 
possesses a large global chiral symmetry, $SU(8)_L \times SU(8)_R$, 
in which the EW and QCD charges are partially embedded.    
 The chiral symmetry as well as the EW symmetry is thus broken by the technifermion-chiral condensate, 
down to the vectorial symmetry $SU(8)_V$, giving rise to 63 NG bosons, the composite technipions, 
where 3 linear combinations of them are eaten by weak bosons $W$ and $Z$, with the rest all becoming very massive
to several-TeV scale by the explicit breaking of the  $SU(8)_L \times SU(8)_R$ symmetry due to the ETC gauge interactions
as well as the EW and the QCD gauge interactions~\footnote{
Although there are only three exact NG bosons to be absorbed into $W$ and $Z$, 
they are linear combinations of all the 
flavors and hence the system is completely different from the  one-doublet model based on 
$SU(2)_L \times SU(2)_R/SU(2)_V$.
Also note that the explicit breaking gauge interactions are rather weak at the relevant energy scale 
so that the masses of all the 60 left-over technipions may be estimated by the perturbative estimate, 
which are actually greatly enhanced  to the several TeV's region by the large anomalous dimension of the walking dynamics. 
The discovery of the technipions at LHC Run II is 
an issue of the one-family model of the walking technicolor~\cite{Kurachi:2014xla}. 
}.

  In addition, the walking technicolor has massive  vector mesons  (technirhos)  which are suggested to have mass much heavier in units of
  the decay constant of technipion $F_\pi$ than that of QCD rho mesons: $M_\rho/F_\pi \gg (M_\rho/F_\pi)_{\rm QCD} \simeq 8$ in various calculations, 
  ladder \cite{Harada:2003dc}, holography \cite{Matsuzaki:2012xx} and the lattice \cite{Appelquist:2014zsa, LatKMI}.
The technirhos should couple to the weak boson pairs 
to be seen as 
distinct resonances in the diboson channel at the LHC. 
Discovering the technirhos in the diboson channel thus would be a smoking-gun of   
the walking technicolor.     In particular, the one-family model predicts a rich spectra of 
63 of such composite vector mesons at several-TeV scale, all of which will be discovered at Run II LHC.

Very recently, the ATLAS collaboration~\cite{Aad:2015owa} has reported an excess of  about 2.5 $\sigma$  global significance at around 2 TeV 
in the diboson channel with the boson-tagged fat dijets. This may imply 
new resonance(s) like the technirhos~\footnote{
The CMS has also done a similar diboson analysis and reported moderate excesses near 
2 TeV~\cite{Khachatryan:2014hpa}.}.

In this paper, we discuss possibilities to explain the excess from the one-family model of the walking technicolor which already successfully explained 
the 125 GeV Higgs in terms of the technidilaton.  
 To address the technihadron phenomenologies at the LHC, 
we employ a low-energy effective theory  for 
the walking technicolor, 
what we call the scale-invariant version of hidden local symmetry (sHLS) model~\cite{Kurachi:2014qma}, 
constructed from the low-lying technihadrons including the technidilaton, 
and 
technipions and technirhos. 
We constrain the couplings of the technirhos to the SM 
fermions and weak gauge bosons.  
by taking into account currently available LHC limits on the technihadrons 
and the size of the resonance width $\lesssim 100$ GeV in the diboson channel with boson-tagged fat dijets  
implied by the ATLAS data~\cite{Aad:2015owa}.

We find that the isospin-triplet technirhos, denoted as $\rho_\Pi^{\pm,3}$, 
can have the couplings large enough to be mainly produced through the Drell-Yan (DY) process analogously to the $\rho$-$\gamma$ mixing 
and predominantly decay into the dibosons. 
We then compute the fat-dijet mass distribution at the almost same level 
of the acceptance and efficiencies for the event selection as those in the ATLAS analysis~\cite{Aad:2015owa},  
and find that the $\rho_\Pi^{\pm, 3}$ with the mass of 2 TeV can account for the currently reported excess 
in the diboson channel.  
The consistency with the electroweak precision test and the discovery possibilities of the  $\rho_\Pi^{\pm,3}$  
in other channels are also discussed.

{\it The model.} -- 
We begin by introducing the effective technihadron Lagrangian (sHLS)
for the one-family model of the walking technicolor~\cite{Kurachi:2014qma}, 
which is a   
version of the original HLS model~\cite{Bando:1984ej} made scale-invariant so as to respect the scale symmetry of the underlying walking technicolor. 
The sHLS Lagrangian is constructed, based on the nonlinear realization of both 
the chiral $SU(8)_L \times SU(8)_R$ symmetry and the scale symmetry, together with 
the hidden local symmetry $[SU(8)_V]_{\rm HLS}$, which is a spontaneously broken gauge symmetry.
The model is characterized by the coset space 
$G/H=[SU(8)_L \times SU(8)_R \times [SU(8)_V]_{\rm HLS}]/SU(8)_V$, with     
the basic quantities being the nonlinear bases $\xi_L, \xi_R$ such that $\xi_L^\dag \xi_R = e^{2i \pi /F_\pi}$ and $\chi(\phi) = e^{\phi/F_\phi}$. They are
parametrized by   
the technipion fields $(\pi)$
and the technidilaton $(\phi)$,  
 with the technipion decay constant $F_\pi$ and the technidilaton decay constant $F_\phi$, respectively. In addition, $\xi_L, \xi_R$ 
are also parameterized by the fictitious NG boson   
fields to be absorbed into the gauge bosons (technirhos) of   
the spontaneously broken HLS,  
$V_\mu$, with the gauge coupling $g$. 
 The practical building blocks are $\hat{\alpha}_{\perp \mu}$ and $\hat{\alpha}_{|| \mu}$, 
which are constructed from the  
$G$-nonlinear bases ($\xi_L, \xi_R$) as 
$\hat{\alpha}_{||, \perp \mu} = (D_\mu \xi_R \xi_R^\dag \pm D_\mu \xi_L \xi_L^\dag)/2i$, 
where the covariant derivatives are $D_\mu \xi_L = \partial_\mu \xi_L - i V_\mu \xi_L + i \xi_L {\cal L}_\mu$ 
and $D_\mu \xi_R = \partial_\mu \xi_R - i V_\mu \xi_R + i \xi_R {\cal R}_\mu$ 
with the external gauge fields $({\cal L}_\mu, {\cal R}_\mu)$ including the SM gauge fields. 
(More details 
such as the transformation properties of these nonlinear bases under the $G$-symmetry and scale symmetry 
are given in Ref.~\cite{Kurachi:2014qma}.)
 
  The Lagrangian corresponding to the $G$- and scale-invariant action 
at the leading order of the derivative expansion is thus written as follows:  
\begin{eqnarray} 
 {\cal L} 
&=& 
F_\pi^2 \chi^2  {\rm tr}[\hat{\alpha}_{\perp \mu }^2] + a F_\pi^2 \chi^2
{\rm tr}[ \hat{\alpha}_{|| \mu}^2 ] 
\nonumber \\ 
&& - \frac{1}{2g^2} {\rm tr}[V_{\mu\nu}^2] 
+ \frac{1}{2} F_\phi^2 (\partial_\mu \chi)^2 
+ \cdots\,, 
 \label{sHLS:Lag:p2}
\end{eqnarray}
where $V_{\mu\nu} = \partial_\mu V_\nu - \partial_\nu V_\mu - i [V_\mu, V_\nu]$ and $a$ is a parameter to be fixed later. 
In Eq.(\ref{sHLS:Lag:p2}) we have indicated, by the ellipsis,  
other terms explicitly breaking the $G$-symmetry and scale symmetry,   
including the $\phi$-potential fixed by the scale anomaly of the underlying walking technicolor~\cite{Matsuzaki:2012vc} 
and the SM gauge boson-kinetic terms.   There are also higher derivative terms of ${\cal O} (p^4)$, etc., which will be discussed later.

The couplings of the technihadrons are obtained by expanding the sHLS Lagrangian 
in powers of the technihadron fields ($\pi, \phi, \rho_\mu$) embedded into $\xi_{L,R}, \chi$ and $V_\mu=g \rho_\mu$.  
The way of embedding technihadron fields and the SM gauge bosons into the 8-flavor matrix forms and 
the explicit expressions of the technihadron couplings (in the unitary gauge of the HLS) 
are given in Ref.\cite{Kurachi:2014qma}. 
Here we just pick up the couplings of 
the isospin-triplet color-singlet technirhos $(\rho_\Pi^\pm, \rho_\Pi^3)$ from the reference, 
which are relevant to the diboson resonance at the mass 
$M_{\rho_\Pi^\pm} = M_{\rho_\Pi^3} \equiv M_{\rho_\Pi} \simeq 2$ TeV~\footnote{The Lagrangian Eq.(\ref{sHLS:Lag:p2}) contains 
other isospin-triplet color-singlet technirhos $\rho_P^{\pm, 3}$ which are not produced by the Drell-Yan process by the $SU(8)$ symmetry \cite{Kurachi:2014qma}. 
Although the 
$SU(8)_V$ symmetry as well as 
the full chiral symmetry $SU(8)_L\times SU(8)_R$  is already violated explicitly in Eq.(\ref{sHLS:Lag:p2})  
due to  the introduction of the SM gauge couplings, 
there still exist  no mixings between the $\rho_\Pi^i$ and $\rho_P^i$ at tree level, thanks to the   spontaneously broken gauge symmetry $[SU(8)_V]_{\rm HLS}$ which is  of course exact without explicit breaking, 
as explicitly shown in Ref.~\cite{Kurachi:2014qma}. 
They do not mix at one-loop level ${\cal O}(p^4) $, neither. 
They can actually mix each other through the electroweak interactions at two-loop order ${\cal O}(p^6)$, 
which are highly suppressed by the loop factor to be negligible for the discussions in the present case. 
}.

The $\rho_\Pi^{\pm, 3}$ couplings to the SM-$f$ fermion pair arise from the mixing between 
the SM gauge bosons and $\rho_\Pi^{\pm, 3}$ present in the $(a F_\pi^2)$ term of Eq.(\ref{sHLS:Lag:p2})  as an analogue of the $\rho$-$\gamma$ mixing in QCD.  
The interaction terms take the form 
\begin{eqnarray} 
&& 
{\cal L}_{\rho_\Pi ff} 
= 
- \sqrt{N_D} \frac{F_\rho}{M_{\rho_{\Pi}}} \Bigg[ 
 e J_\mu^{\rm em} \rho^{3\mu}_\Pi  
\nonumber \\ 
&& 
+ \frac{e(c^2-s^2)}{2 sc}\frac{1}{1 - m_Z^2/M_{\rho_\Pi}^2} J_\mu^Z \rho^{3\mu}_\Pi   
\nonumber \\ 
&& 
+ \frac{e}{2 s}\frac{1}{1 - m_W^2/M_{\rho_\Pi}^2} (J_\mu^{W^+} \rho^{+ \mu}_\Pi + {\rm h.c.}  )  
\Bigg] 
\,, \label{rho-ff}
\end{eqnarray}
where $e$ and $s$ ($c^2=1-s^2$) respectively denote the electromagnetic coupling and the (sine of) standard weak mixing angle.   
The SM fermion currents are defined as 
$  J_\mu^{\rm em} = e \sum_f \bar{f} \gamma_\mu Q_{\rm em}^f f $, 
$  J_\mu^Z = e/(sc) \sum_f \left[ \bar{f}_L \gamma_\mu (\tau^f_3 - s^2 Q_{\rm em}^f) f_L 
+ \bar{f}_R \gamma_\mu (- s^2 Q_{\rm em}^f) f_R \right] 
$ and 
$ J_\mu^{W^+} = e/(\sqrt{2}s) \sum_f \bar{f}_{u L} \gamma_\mu f_{d L}  $ ($J_\mu^{W^-}=(J_{\mu}^{W^+})^\dag$)
with the electromagnetic charge for the $f$-fermion $Q_{\rm em}^f$ and the isospin charges  
$\tau^f_3 = \pm 1/2$ for the up- and down-sector $f_{u,d}$-fermions.   
The $\rho_\Pi$-SM gauge boson-mixing strength $F_\rho$ is expressed in terms of the sHLS parameters as 
$F_\rho = \sqrt{a} F_\pi$.  
The prefactor $\sqrt{N_D}$ stands for the number of electroweak doublets formed by technifermions, which is 4 in the case of the one-family model.   
Note that the $N_D$ dependence is canceled out in the combination $(\sqrt{N_D} F_\rho)$: 
$\sqrt{N_D} F_\rho = \sqrt{N_D} \sqrt{a} F_\pi = \sqrt{a} v_{\rm EW}$ where $v_{\rm EW} \simeq 246$ GeV (See also Eq.(\ref{Fpi}) below).   
In reaching Eq.(\ref{rho-ff}) we have also used the $\rho_\Pi$ mass formula obtained from the ($a F_\pi^2$) term of Eq.(\ref{sHLS:Lag:p2}): 
\begin{equation} 
 M_{\rho_\Pi} = \sqrt{a} g F_\pi = 2\,  {\rm TeV}
 \,, \label{mass:formula}
\end{equation}
 where we have set it to 2 TeV for the present analysis.

As to the couplings to the weak gauge bosons, 
the $\rho_\Pi$ mainly couples to the longitudinal modes 
$(W_\mu, Z_\mu)_L = (\partial_\mu \pi_W/m_W, \partial_\mu \pi_Z/m_Z)$~\footnote{
The $\rho_\Pi$ couplings to the transverse modes of the 
weak gauge bosons turn out to be highly suppressed by a factor of $(m_{W/Z}/M_{\rho_\Pi})^2$, 
hence the $\rho_\Pi$ are hardly produced via the vector boson fusion and vector-boson associate processes.     
}, where $\pi_W$ and $\pi_Z$ stand for the NG bosons eaten by the $W$ and $Z$, respectively. 
The $\rho_\Pi$-$W_L$-$W_L/Z_L$ vertices thus arise from 
the $(a F_\pi^2)$ term of Eq.(\ref{sHLS:Lag:p2}):  
\begin{eqnarray} 
&& {\cal L}_{\rho_\Pi W_LW_L/W_LZ_L}  
= 
\frac{1}{\sqrt{N_D}} g_{\rho\pi\pi} i 
\Bigg[
\partial^\mu \pi_W^+ \pi_W^- \rho_{\Pi \mu}^3 
\nonumber \\ 
&& + 
(\partial^\mu \pi_W^- \pi_Z - \partial^\mu \pi_Z \pi_W^-) \rho_{\Pi \mu}^+ 
\Bigg] 
+ {\rm h.c.} 
\,, 
\label{rho-pi-pi}
\end{eqnarray}
  where 
$g_{\rho \pi\pi} = (1/2) ag$ in terms of the original Lagrangian parameters. 
Note the prefactor $(1/\sqrt{N_D})$ in front of the $\rho$-$\pi$-$\pi$ coupling:  
this is the characteristic feature of the one-family model of walking technicolor 
and realizes the smaller $\rho \to W_LW_L$ width by a factor of $(1/N_D)=1/4$,  
compared to the naive scale-up version of QCD having only the one weak doublet, 
as will be seen clearly below.

It turns out that the $\rho_\Pi$ does not couples to the technidilaton $\phi$ involving 
the SM gauge bosons ($W, Z$ and photon $\gamma$) because of the scale invariance~\cite{Fukano:2015uga}.

 As we mentioned before, all the 60 technipions not eaten by the $W$ and $Z$ bosons in the one-family model acquire  
masses   
due to the explicit breaking of the $SU(8)_L\times SU(8)_R$ chiral symmetry by the EW and QCD as well as the ETC gauge couplings,
which are enormously enhanced to  the order of ${\cal O}(\textrm{a few TeV})$ by the large anomalous dimension $\gamma_m \simeq 1$
as a salient  feature of the walking technicolor~\cite{Kurachi:2014xla}, and hence should be  
larger than the technirho mass. 
Thereby, we will ignore the 2 TeV $\rho_\Pi$ couplings to technipions.

In addition to the leading order terms in Eq.(\ref{sHLS:Lag:p2}), 
one may incorporate higher-derivative coupling terms, so-called ${\cal O}(p^4)$ terms, 
introduced as in the original HLS formulation applied to QCD~\cite{Tanabashi:1993np,Harada:2003jx}, 
\begin{eqnarray} 
  {\cal L}' 
&=&  
z_3 {\rm tr}[\hat{\cal V}_{\mu\nu} V^{\mu\nu}] 
+ i z_4 {\rm tr}[V^{\mu\nu} \hat{\alpha}_{\perp \mu}\hat{\alpha}_{\perp \nu}]  
\,, \label{z3-z4}
\end{eqnarray}
where $\hat{\cal V}_{\mu\nu}$ becomes $1/2 [\partial_\mu ({\cal R}_\nu + {\cal L}_\nu) - \partial_\nu ({\cal R}_\mu + {\cal L}_\mu)] + $ 
(non-Abelian terms and technipion terms) in the unitary gauge of the HLS.    
The order of magnitude for the parameters $z_3$ and $z_4$ can be estimated 
by the naive dimensional analysis to be $z_{3,4}={\cal O}(N_C/(4\pi)^2)={\cal O}(10^{-2})$ for $N_C=4$.   
These terms affect the $\rho_\Pi$-$f$-$f$ and $\rho_\Pi$-$W_L$-$W_L/Z_L$ couplings at the on-shell of the $\rho_\Pi$: 
the $F_\rho$ coupling in Eq.(\ref{rho-ff}) and the $g_{\rho\pi\pi}$ coupling in Eq.(\ref{rho-pi-pi}) are then modified as 
\begin{eqnarray} 
F_\rho = \sqrt{a} F_\pi 
&\rightarrow& F_\rho = \sqrt{a} F_\pi (1 - g^2 z_3) 
\,, \nonumber \\ 
g_{\rho\pi\pi} = \frac{1}{2} ag 
&\rightarrow& 
g_{\rho\pi\pi} = \frac{1}{2} ag (1 - \frac{1}{2} g^2 z_4) 
\,. \label{modify}
\end{eqnarray}

Thus, by including the possible higher derivative terms, 
the couplings relevant to the 2 TeV $\rho_\Pi$ phenomenology are found to be 
controlled by the four parameters $F_\pi$, $g$, $z_3$ and $z_4$ with  the parameters $a$ and $F_\phi$ being redundant for the present  analysis.  
In place of the original parameters $z_3$ and $z_4$, we shall use 
$F_\rho$ and $g_{\rho\pi\pi}$ with the replacement in Eq.(\ref{modify}) 
taken into account. 
Among those four parameters, 
the technipion decay constant $F_\pi$ is related to the EW scale $v_{\rm EW}\simeq 246$ GeV for the one-family model with four weak-doublets, $N_D=N_F/2=4$, as
\begin{equation} 
F_\pi = v_{\rm EW}/\sqrt{N_D}\simeq 123 \,  {\rm GeV} 
\,,  \label{Fpi}
\end{equation} 
through the $W/Z$ mass formula obtained by examining the $F_\pi^2$ term in Eq.(\ref{sHLS:Lag:p2}). 
The HLS gauge coupling $g$ is then determined through the $\rho_\Pi$ mass formula in Eq.(\ref{mass:formula}), once 
the $\rho_\Pi$ mass is set to 2 TeV and the parameter $a$ is chosen. 
Then, with two inputs $(F_\pi,\, M_{\rho_\Pi})$, we are left only with the two parameters 
\begin{equation}
\left( F_\pi, \, g, \,z_3,\, z_4\right)
\longrightarrow \left(F_\rho, \, 
g_{\rho\pi\pi} \right), 
 \end{equation} 
which control the $\rho_\Pi$ couplings to the SM fermions 
(Eq.(\ref{rho-ff})) and the weak gauge bosons (Eq.(\ref{rho-pi-pi})), respectively.

{\it Constraining the $\rho_\Pi$ couplings.} -- 
Using Eqs.(\ref{rho-ff}), (\ref{rho-pi-pi}) we compute 
the partial decay widths of the $\rho_\Pi^{\pm, 3}$ as functions of $F_\rho$ and $g_{\rho\pi\pi}$. 
The $\rho^{\pm, 3}_\Pi$ couplings  are implemented 
by using the \texttt{FeynRules}~\cite{Christensen:2008py} 
and the 
\texttt{MadGraph5\_aMC@NLO}~\cite{Alwall:2014hca} is used for the estimation of the decay widths. 
  We constrain the size of the total widths to be $\lesssim$ 100 GeV 
in light of the ATLAS data on the diboson-tagged dijet mass distribution~\cite{Aad:2015owa}, 
so that the relevant couplings $F_\rho$ and $g_{\rho\pi\pi}$ are constrained.

\begin{figure}[ht]
\begin{center}
   \includegraphics[width=7.0cm]{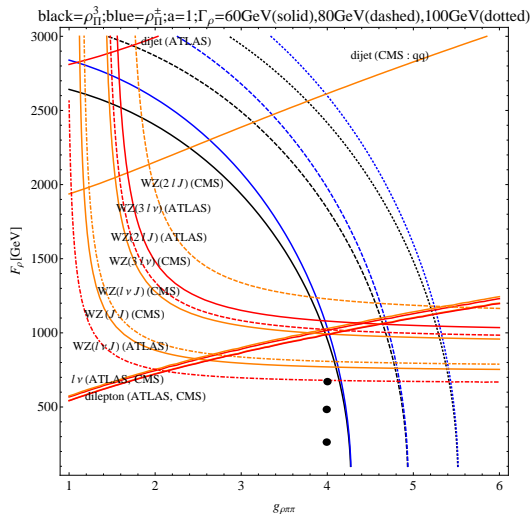} 
\caption{ 
The contours in the $(g_{\rho\pi\pi}, F_\rho)$ plane drawn by requiring the total widths 
$\Gamma_{\rho_\Pi^3}$ (black curves, in the right-middle) or $\Gamma_{\rho^\pm_\Pi}$ (blue curves, in the left-middle) 
to be  60 (solid), 80 (dashed), 100 (dotted) GeV at the mass of 2 TeV. 
The 95\% C.L. upper limits at around 2 TeV 
from the LHC experiments~\cite{Khachatryan:2015sja,Khachatryan:2014xja,Aad:2014pha,Aad:2014xka,Khachatryan:2014gha,Aad:2015ufa,Khachatryan:2014tva,ATLAS:2014wra,Aad:2014cka,Khachatryan:2014fba,Khachatryan:2014hpa} have also been shown together. 
The regions above those curves are excluded. 
The there black blobs denote the reference points in Eq.(\ref{Frho-value}) with Eq(\ref{grhopipi-value}).    
\label{Frho-grhopipi-Cons-contours-a1}
}
\end{center} 
 \end{figure}

In Fig.~\ref{Frho-grhopipi-Cons-contours-a1} we show the contour plot in the $(F_\rho, g_{\rho\pi\pi})$ plane 
together with 
the 95\% C.L. limits at around 2 TeV  
for $W'/Z'$ candidates reported from the ATLAS and CMS experiments~\cite{Aad:2014aqa,Khachatryan:2015sja,Khachatryan:2014xja,Aad:2014pha,Aad:2014xka,Khachatryan:2014gha,Aad:2015ufa,Khachatryan:2014tva,ATLAS:2014wra,Aad:2014cka,Khachatryan:2014fba,Khachatryan:2014hpa}~\footnote{
In addition to limits shown in Fig.~\ref{Frho-grhopipi-Cons-contours-a1}, 
there are other limits from $W'/Z' \to $ Higgs plus weak bosons reported from the ATLAS and CMS~\cite{Aad:2015yza,Khachatryan:2015bma}, 
in which the analyses are based on the Higgs decay to $bb$ or $WW$. 
However, the $\rho_\Pi$s in the present study 
do not decay to the Higgs candidate (technidilaton)~\cite{Fukano:2015uga},   
so we have not incorporated those constraints in the figure.  
}. 
The upper bounds of the cross sections used to make the plots 
are as follows: 
\begin{eqnarray} 
&& \sigma_{WZ(3 l \nu)}^{\rm ATLAS}[{\rm fb}] 
\le 22 \,, \qquad 
\sigma_{WZ(3 l \nu)}^{\rm CMS}[{\rm fb}] 
\le 19 \,, \nonumber \\ 
&& \sigma_{WZ(2 l J)}^{\rm ATLAS}[{\rm fb}] 
\le 20 \,, \qquad 
\sigma_{WZ(2 l J)}^{\rm CMS}[{\rm fb}] 
\le 27 \,, \nonumber \\
&& \sigma_{WZ(l \nu J)}^{\rm ATLAS}[{\rm fb}] 
\le 9.5 \,, \qquad 
\sigma_{WZ(l \nu J)}^{\rm CMS}[{\rm fb}] 
\le 13
\, \nonumber \\ 
&& \sigma_{WZ(JJ)}^{\rm CMS}[{\rm fb}] 
\le 12 \,, 
\nonumber \\ 
&& \sigma_{l \nu}^{\rm ATLAS}[{\rm fb}] 
\le 0.41 \,, \qquad 
\sigma_{l \nu}^{\rm CMS}[{\rm fb}] 
\le 0.42
\, \nonumber \\ 
&& \sigma_{2l}^{\rm ATLAS}[{\rm fb}] 
\le 0.24 \,, \qquad 
\sigma_{2l}^{\rm CMS}[{\rm fb}] 
\le 0.25
\, \nonumber \\ 
&& \sigma_{2j}^{\rm ATLAS}[{\rm fb}] 
\le 130 \,, \qquad 
\sigma_{2j(qq)}^{\rm CMS}[{\rm fb}] 
\le 58
\,, 
\end{eqnarray}
where in the last line 
we have quoted the upper limit set on generic narrow resonances (with the width being 0.1 percent of the mass)
decaying to the $qq$-jet reported from 
the CMS group~\cite{Khachatryan:2015sja}, while the ATLAS bound~\cite{Aad:2014aqa} includes all the jet candidates.    
The $\rho_\Pi$ cross sections have been computed by using 
\texttt{CTEQ6L1}~\cite{Stump:2003yu} parton distribution functions. 
Though the result is not changed by the value of the parameter $a$~\footnote{
The parameter choice a = 1 is special, corresponding to the
locality of the deconstructed extra dimension (3-site model in
the linear moose.
}, for just a practical reason, we have chosen $a=1$ 
in computing the cross sections and decay widths. 
 From the figure, we see that  
 the $g_{\rho \pi\pi}$ coupling is constrained by requiring the total width for $\rho_\Pi^\pm$ to be 
$\Gamma_{\rho_\Pi^\pm} \lesssim 100$ GeV as $g_{\rho\pi\pi} \lesssim 5.5$. 
Hereafter, as a reference value, we shall take~\footnote{
The $g_{\rho\pi\pi}$ is slightly smaller than the QCD value $\simeq 6$ for $N_C=3$, which can be 
understood by the $N_C$ scaling, $g_{\rho\pi\pi}|_{N_C=4} \sim \sqrt{3/N_C} g_{\rho\pi\pi}|_{N_C=3}$ 
and destructive axialvector-$a_1$ contribution  which is mimicked by the size of 
$z_4$ in the sHLS model (see also summary and  discussion in the later section). }    
 \begin{equation} 
  g_{\rho \pi\pi} = 4  
  \, . \label{grhopipi-value}
 \end{equation}  
 As to the $F_\rho$, we may choose a typical set of the values satisfying 
the current LHC limits as 
\begin{equation} 
  F_\rho [{\rm GeV}] = 250\,,  500\,, 650  
  \,, \label{Frho-value} 
\end{equation} 
in which the first value (250 GeV) is supported from nonperturbative calculations of a large $N_F$ walking gauge theory~\cite{Harada:2003dc,Appelquist:2014zsa},  
which are currently possible to quote at hand, 
the third one (650 GeV) is a representative of the maximal value satisfying the dilepton constraint, 
and the middle value (500 GeV) is just a sample in between the other two~\footnote{ 
 In terms of the original Lagrangian parameters, $g$, $z_3 $ and $z_4$ in Eqs.(\ref{sHLS:Lag:p2}), (\ref{z3-z4}) and (\ref{modify}), 
we have $g\simeq 16$, $z_4 \simeq 4.0 \times 10^{-3}$ and $z_3 \simeq (- 4.0, - 12, - 18) \times 10^{-3}$ 
for $a=1$, $g_{\rho\pi\pi}=4$, $M_{\rho_\Pi}=2$ TeV and $F_\rho = (250, 500, 650)$ GeV.  
The estimated size of $z_3$ and $z_4$ are compatible 
with the order of magnitude estimated by the naive dimensional analysis, ${\cal O}(N_C/(4\pi)^2)$ (see also text). 
The large value of $g$
implies large corrections from higher order in the HLS chiral perturbation theory, with the expansion parameter $\xi= N_F\cdot g^2/(4\pi)^2\simeq N_F \cdot p^2 /(4\pi F_\pi)^2|_{p=M_\rho} \simeq 16 \gg 1\, (N_F=8)$, in comparison with the real-life QCD ($N_F=3$), $\xi \simeq 1$.
Origin of the large corrections will be discussed in the Summary and discussion. 
}.

 For the values in Eqs.(\ref{grhopipi-value}) and (\ref{Frho-value}), 
we have the total widths,   
\begin{eqnarray} 
&& \Gamma_{\rho_\Pi^3} \, [{\rm GeV}]  \simeq (53, 55, 56)
\,, \nonumber \\ 
&& \Gamma_{\rho_\Pi^\pm} \, [{\rm GeV}] \simeq (53, 54, 55)
\,, 
\nonumber \\ 
&& 
{\rm for} \quad F_\rho [ {\rm GeV} ]= (250, 500, 650)  
\,, 
\end{eqnarray} 
and the dominant branching ratios,  
\begin{eqnarray} 
&& {\rm Br}(\rho_\Pi^3 \to WW) \, [\%]  
\simeq (99, 96, 94) 
\,, \nonumber \\ 
&& 
{\rm Br}(\rho_\Pi^\pm \to W^\pm Z) \, [\%] 
\simeq (99, 97, 95)
\,, \nonumber \\  
&& {\rm for} \quad F_\rho [ {\rm GeV} ]= (250, 500, 650) 
\,.  
\end{eqnarray}

  The $\rho_\Pi^{\pm, 3}$ are mainly produced via the DY process with the $F_\rho$ as in Eq.(\ref{Frho-value}).  
The total cross sections $\sigma_{\rm DY}(pp \to \rho_\Pi^{\pm, 3})$ at the center of mass energy $\sqrt{s}=$ 8 TeV are computed 
by using the \texttt{CTEQ6L1}~\cite{Stump:2003yu} parton distribution functions to be 
\begin{eqnarray} 
 && \sigma_{\rm DY}(pp \to \rho_\Pi^{3}) \, [{\rm fb}]
 \simeq (0.7, 2.8, 4.7) 
 \,, \nonumber\\ 
  && \sigma_{\rm DY}(pp \to \rho_\Pi^{\pm}) \, [{\rm fb}]
 \simeq (1.4, 5.4, 9.3)  
 \,, \nonumber\\ 
 && {\rm for} \quad F_\rho [ {\rm GeV} ]= (250, 500, 650) 
 \,. 
\end{eqnarray}  
Other subdominant decay and production properties 
are to be given in another publication.

  \begin{figure}[ht]
\begin{center}
   \includegraphics[width=7.5cm]{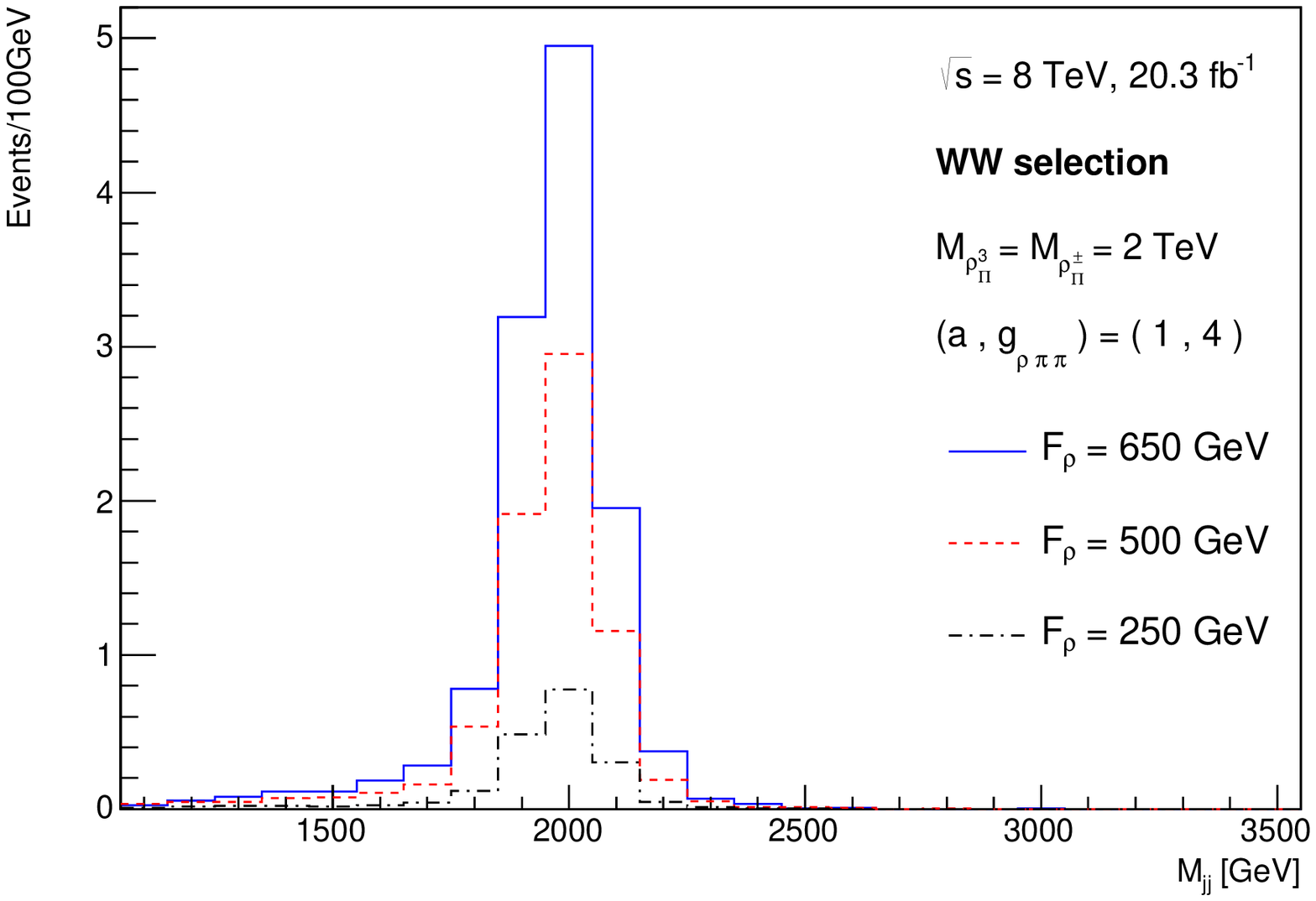} 
   \includegraphics[width=7.5cm]{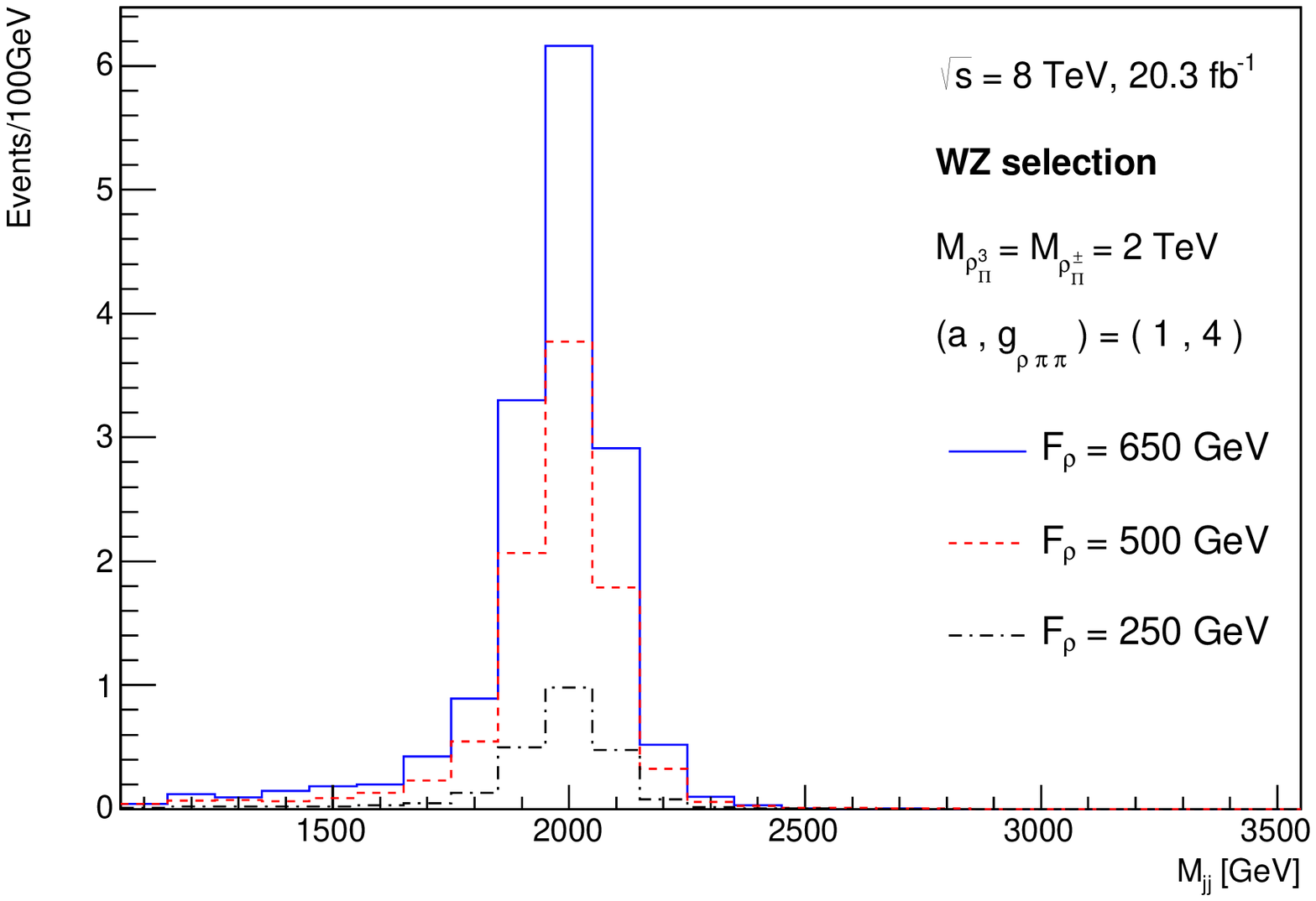} 
\caption{ 
The fat-dijet mass distribution $M_{JJ}$ of $\rho_\Pi^{\pm,3}$ for the $WW$ 
(top panel) and the $WZ$ (bottom panel) selections 
 at 2 TeV  with $F_\rho[{\rm GeV}]=250$ (dashed-dotted line), 500 (dashed line), 650 (solid line). 
\label{cut7-MJJ}
}
\end{center} 
 \end{figure}

{\it The technirho signals at 2 TeV in the 
fat-dijet mass distribution with the diboson-tagged.} --   
Now that all the parameters are fixed, we are ready to discuss 
the $\rho_\Pi$ signals in the dijet mass distribution with  
the EW diboson tagged, in comparison with 
the ATLAS data~\cite{Aad:2015owa}. 
To calculate the dijet mass distribution, 
we use the \texttt{FeynRules}~\cite{Christensen:2008py} to implement the $\rho_\Pi$ couplings, fixed as above,  
into the \texttt{MadGraph5\_aMC@NLO}~\cite{Alwall:2014hca} and generate the cross sections at the 
parton level. 
For the hadronization and parton showering, 
\texttt{PYTHIA 8.1.86}~\cite{Sjostrand:2006za} is used. 
The events are generated by using the \texttt{CTEQ6L1}~\cite{Stump:2003yu} parton distribution functions 
and the jets are reconstructed through the Cambridge-Aachen algorithm~\cite{Dokshitzer:1997in,Wobisch:1998wt} 
with the radius parameter $R=1.2$ (C/A $R=1.2$)
by the \texttt{FastJet 3.0.6}~\cite{Cacciari:2011ma}.

To make a direct comparison with the ATLAS analysis, the C/A $R=1.2$ jets (called fat-jets hereafter) are processed through a splitting and filtering algorithm 
described as ``BDRS-A" in Ref.~\cite{BDRS-A}, similar to the algorithm in Ref.~\cite{BDRS} but modified for 
high-$p_T$ boson jets in ATLAS analysis. After that, we apply the same event selections as in Ref.~\cite{Aad:2015owa}:  
i) the number of fat-jets $\ge 2$;   
ii) the momentum balance $\sqrt{y_f} = \min(p_T^{j1},p_T^{j2})\Delta R_{12}/m_{12} \ge 0.45$, where $p_T^{j1}$ and $p_T^{j2}$ are the transverse 
momenta of the two leading subjets found by the BDRS-A algorithm, $\Delta R_{12}$ and $m_{12}$ are the $\eta$-$\phi$ distance and mass of the two subjets, respectively;
iii) the transverse momentum $p_T$ for the leading fat-jet $J_1$, $p_T(J_1) \ge 540$ GeV;   
iv) the pseudo-rapidity $\eta$ for the two leading fat-jets $J_{1,2}$, $|\eta(J_1,J_2)| \le 2$;  
v) the rapidity difference $\Delta y$ between the two leading fat-jets, $ \Delta y = |y(J_1) - y(J_2)  | \le 1.2$;   
vi) the $p_T$ asymmetry for the two leading fat-jets, $\frac{p_T(J_1)-p_T(J_2)}{p_T(J_1) + p_T(J_2)} \le 0.15$;  
vii) the number of charged-particle tracks associated with the original ungroomed fat-jet, $n_{\rm tr} < 30$;   
viii) the range of the fat-jet masses, $M_{J_1,J_2}$, 
$82.4-13 \le M_{J_1,J_2} \, [{\rm GeV}] \le 82.4 + 13$ for the $WW$ selection, 
 $92.8-13 \le M_{J_1,J_2} \, [{\rm GeV}] \le 92.8 + 13$ for the $ZZ$ selection, 
and  $92.8-13 \le M_{J_1} \, [{\rm GeV}] \le 92.8 + 13$, 
$82.4-13 \le M_{J_2} \, [{\rm GeV}] \le 82.4 + 13$, 
for the $WZ$ selection where $M_{J_1} > M_{J_2}$.  
For vii), since the charged-particle track reconstruction is not performed in this study, the $n_{\rm tr}$ requirement is not directly applied. 
Instead, assuming that the $\rho_\Pi^{\pm,3}$ has the same $n_{\rm tr}$ distribution as the $W'$ signal used in ATLAS (Fig.~1b of Ref.~\cite{Aad:2015owa})
and hence the scaling factor ($0.90\pm0.08$ in Ref.~\cite{Aad:2015owa}) defined as the ratio of the $n_{\rm tr}$ cut efficiency in data 
to that in the $W'$ simulation is applicable, 
the $\rho_\Pi^{\pm,3}$ signal yields are scaled by the square of the product of the cut efficiency and the scaling factor. 
The cut efficiency is estimated from the $W'$ signal distribution shown in Fig.~1b of Ref.~\cite{Aad:2015owa}. 
In addition, to account for the migration due to ATLAS detector resolution, the fat-jet momentum/energy and mass values are smeared by Gaussian distributions 
with the mean of 0 and the standard deviations of 5\% (as in Ref.~\cite{Aad:2015owa}) and 8\% (taken from $600<p_T<1000$~GeV bin 
in Table~2 of Ref.~\cite{BDRS-A}), respectively.

In Fig.~\ref{cut7-MJJ} we show the technirho $\rho_\Pi$ signals in the 
fat-dijet mass distributions with the diboson tagged 
for the $WW$ (top panel) and $WZ$ (bottom panel) selections 
at $\sqrt{s}=8$ TeV and the integrated luminosity ${\cal L}=20.3 \, {\rm fb}^{-1}$. 
  From the ATLAS result~\cite{Aad:2015owa} we see that at 2.0 (1.9) TeV,   
  the number of observed events are 8(5) and 7(4) per 100 GeV bin over the expected background $\simeq$ 2 (3) 
  for the $WZ$ and $WW$ selections, respectively.  
  In comparison with the data,  Fig.~\ref{cut7-MJJ} implies that 
the 2 TeV $\rho_\Pi^{\rm \pm, 3}$ with $F_\rho = 650$ GeV  can explain 
the excess with about 3 $\sigma$ local (2.5 $\sigma$ global) significance  
for both the $WW$ and $WZ$ selections.

  \begin{figure}[ht]
\begin{center}
   \includegraphics[width=7.5cm]{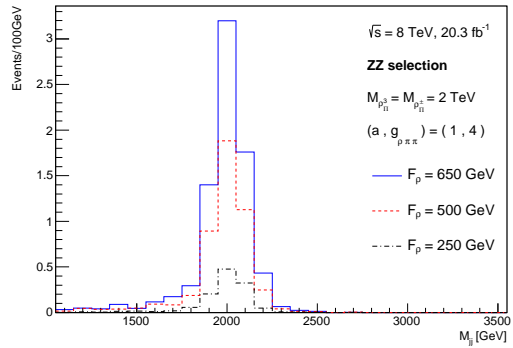} 
\caption{ 
The fat-dijet mass distribution $M_{JJ}$ of $\rho_\Pi^{\pm,3}$ for the $ZZ$ selection 
 at 2 TeV  with $F_\rho[{\rm GeV}]=250$ (dashed-dotted line), 500 (dashed line), 650 (solid line).  
\label{cut7-MJJ-ZZ}
}
\end{center} 
 \end{figure}

The ATLAS collaboration~\cite{Aad:2015owa} have also reported 
an excess by about 2.9 $\sigma$ (local significance) in the $ZZ$ channel. 
As listed in the item viii) of the event selection above, 
the $WW, WZ$ and $ZZ$ selections are distinguished only by 
the fat-jet mass ranges, hence the $Z$ boson (e.g. its charge) 
is not clearly identified in the analysis. 
Therefore, the $\rho_\Pi \to WW/WZ$ events may contaminate in the $ZZ$ channel, 
 although the $\rho_\Pi$ does not couple to $ZZ$ in the model discussed here.  
In Fig.~\ref{cut7-MJJ-ZZ} we plot the fat-jet mass distributions of the 2 TeV 
$\rho_\Pi^{\pm,3}$ in the $ZZ$ selection for $F_\rho[{\rm GeV}]=250, 500, 650$. 
  From the ATLAS result~\cite{Aad:2015owa} we read off 
the 3 (5) observed events/100 GeV over the expected background $\simeq 0.6$ $(0.9)$ at 2.0 (1.9) TeV.  
The figure shows that the 2 TeV $\rho_\Pi^{\pm, 3}$ with $F_\rho=650$ GeV 
can also account for the excess in the $ZZ$ channel.

{\it Summary and discussion.} --  
In summary, 
we have discussed the 2 TeV diboson signals for the  
isospin-triplet color-singlet technirhos ($\rho_{\Pi}^{\pm, 3}$) 
in the one-family model of the walking technicolor. 
It has been shown that the $\rho_{\Pi}^{\pm, 3}$ with the narrow width $\lesssim$ 100 GeV and 
a large Drell-Yan coupling, 
such as $F_\rho/M_{\rho_\Pi} (= 650\,{\rm GeV}/2000\,{\rm GeV}) \simeq 0.3$,   
can account for the excesses in the fat-dijet mass distributions in the diboson channel    
recently reported from the ATLAS group, consistently with the current-observed limits in  other channels.

Some comments are in order:


1) We have shown in Fig.~\ref{cut7-MJJ-ZZ} that the 2 TeV $\rho_\Pi^{\pm, 3}$ with $F_\rho=650$ GeV 
is consistent with the excess seen in the $ZZ$ channel. This is largely because the $\rho_\Pi \to WW/WZ$ events 
become broadly distributed in fat-jet mass due to finite detector resolution and hence some fraction of events 
are selected by the fat-jet mass cut of the $ZZ$ selection. The experimental capability to separate $WW/WZ$ 
from $ZZ$ in decays of high-mass resonances will be an important ingredient to further test the models discussed 
in this study.  
Exploiting the leptonic $ZZ$ decays ($\ell\ell q\bar{q}$, $\ell\ell \ell'\ell'$) will provide a handle to separate 
them from $WW$ or $WZ$ decay, while it requires a large dataset due to a small $Z \to \ell\ell$ branching fraction. 
Hence it is interesting to explore separation techniques based on the hadronic $W/Z$ decays.

Shown in Fig.~1 of Ref.~\cite{Aad:2015owa} are reconstructed fat-jet mass distributions of 
hadronically decaying $W$ and $Z$ bosons, as simulated for the 1.8~TeV graviton decaying into $WW/ZZ$. 
As clearly seen, the $W$ and $Z$ bosons can be statistically separated by $\sim 10$ GeV at peak positions 
if the excess remains and sufficient statistics is accumulated. 
In addition, the ATLAS study on charged-particle tracks associated with fat-jets from high-$p_T$ hadronic $W^\pm$ 
bosons Ref.~\cite{TheATLAScollaboration:2013sia} shows that the $W^+$ and $W^-$ can be separated at 
the level of $\sim 50$\% $W^+$ efficiency with $W^-$ rejection of about 4. This technique is expected to be less performant
for separation between $WW$ and $ZZ$ resonances because a mixture of $W^\pm$ in $WW$ would be 
less distinguishable from $Z$ in $ZZ$ on average. We however expect this charge measurement to provide additional 
discrimination between $WZ$ and $ZZ$ resonances by adopting the technique separately to higher and lower mass 
fat-jets in selected events.


2) The sHLS model we have employed throughout the present study can be viewed as 
the low-energy effective theory induced from the underlying walking technicolor.  
Actually, the DY coupling of the technirho, $(F_\rho/M_\rho)$, detected at the LHC, 
should include contributions from not only the walking technicolor sector, but also an ETC 
sector, which communicates between the technifermion and the SM fermion sectors and hence necessary to account for 
the SM-fermion mass generation. The ETC yields an effective four-fermion interaction among the technifermions of the form $(\bar F \gamma_\mu T_a F)^2$, 
with $T_a$ being the $SU(N_F)$ generators, 
which affects $F_\rho$.
In that sense, the size of the $F_\rho$ constrained by the current LHC data, mainly from the dilepton channel, 
would imply the desired amount of the ETC 
contributions: 
one may take the value of the DY coupling estimated only from the walking technicolor sector 
to be $F_\rho^{\rm TC} \simeq 250$ GeV, supported from the result of nonperturbative calculations presently at hand~\cite{Harada:2003dc,Appelquist:2014zsa}. 
Then the rest may be supplied from the ETC, say $F_\rho^{\rm ETC}[{\rm GeV}] =(0, 250, 400)$ for 
the total $F_\rho [{\rm GeV}] = ( F_{\rho}^{\rm TC} + F_{\rho}^{\rm ETC}) [{\rm GeV}] = (250, 500, 650)$. 
This implies an indirect constraint on modeling of the ETC  
derived from the current LHC data from  
the dilepton channel. 
Note that in the present analysis, 
such ETC effects can be, in a sense, mimicked by the parameter $z_3$ in Eq.(\ref{z3-z4}) which shifts 
the $F_\rho$ as in Eq.(\ref{modify}) just like $F_\rho \to F_\rho = F_\rho^{\rm TC} + F_\rho^{\rm ETC}$.


3)
With such a large DY coupling of the technirho at hand, one might suspect 
the large contribution to the $S$ parameter~\cite{Peskin:1990zt+}. 
One can in fact estimate the size of the $S$ coming only from the technirho contribution 
within the sHLS model, to find 
$S|_{\rho} = 4 \pi N_D (F_\rho/M_{\rho_\Pi})^2 \simeq {\cal O}(10)$ for $F_\rho = 650$ GeV,   
where $N_D =4$ for the  
one-family model. 
However, the techni-axialvector (techni-$a_1$) contribution may cancel the large $S|_\rho$ term: 
in the one-family model of the walking technicolor, 
it has been suggested from several approaches~\cite{Harada:2003dc,Haba:2010hu,Matsuzaki:2012xx,Appelquist:2014zsa} that 
the masses of the techni-$\rho$ and -$a_1$ mesons are degenerate, $M_{\rho_\Pi} \simeq M_{a_{1\Pi}}$, 
due to the characteristic walking feature. 
Taking into account this, one may add the techni-$a_1$ meson contribution to the $S$ as 
$S = S|_\rho + S|_{a_1} = 4 \pi N_D (F_\rho/M_{\rho_\Pi})^2 [1 - (F_{a_1}/F_{\rho})^2 ]$.        
Thus, if $F_{a_1} \simeq F_{\rho}$ (including possible ETC contributions), 
the $S$ parameter can be vanishingly small to be $\simeq 0$, as it happens in a different context~\cite{Casalbuoni:1995yb}:   
this can actually take place in the one-family model of the walking technicolor, in 
view of 
a holographic dual~\cite{MY}.    If it  is the case we may expect a new type DY process through the techni-$a_1$ meson at 2 TeV.

4) Note also that the presence of the techni-$a_1$ mesons generically modify 
the $g_{\rho\pi\pi}$ coupling due to the mixing between the pions and axialvector mesons, 
as in the case of the generalized HLS model~\cite{Bando:1987ym} in application to QCD. 
Such a role of the shift effects can actually be played by the $z_4$ term in Eq.(\ref{z3-z4}), instead of the techni-$a_1$ mesons,  
as in Eq.(\ref{modify}).


5) 
As to the DY coupling, it is furthermore indicated that if the diboson excess is explained by 
the presence of the 2 TeV technirhos $\rho_{\Pi}^{\pm, 3}$ with the large DY coupling $F_\rho=650$ GeV which barely satisfies 
the current dilepton limit, 
the 2 TeV $\rho_\Pi^{\pm, 3}$, 
as well as the 2 TeV techni-$a_1$ mesons having the same amount of the DY coupling $F_{a_1}$ to suppress the $S$ parameter,   
can necessarily be seen also in the dilepton channel as quite narrow resonances 
at the LHC-Run II, which could be possible at the earlier stage of the running. 
To put it the other way around, if the $\rho_\Pi$ and techni-$a_1$ would have somewhat small DY coupling like $F_\rho=250$ GeV or less 
and the diboson excess goes away in the future, 
it might be hard to detect the $\rho_\Pi$ and techni-$a_1$ at the LHC. 
Other technirhos in the one-family model could then be discovered  through 
other channels.  
In this sense, the fate of the currently reported diboson excess is linked with which 
channels and technivectors and techni-axialvectors 
are accessible at the LHC-Run II.


More detailed analysis on the promising LHC signals for the technivector and axialvector mesons 
in the one-family model of the walking technicolor 
will be pursued in the future.

\acknowledgments 

We are grateful to K.~Tobe for useful comments.  
This work was supported in part by 
the JSPS Grant-in-Aid for Young Scientists (B) \#15K17645 (S.M.) 
and MEXT KAKENHI Grant-in-Aid for Scientific Research on Innovative Areas \#25105011 (M.K.).

\end{document}